\documentclass[
 reprint,superscriptaddress, amsmath,amssymb,
 aps,
floatfix,nofootinbib
]{revtex4-2}

\usepackage{graphicx}
\usepackage{dcolumn}
\usepackage{bm}
\usepackage{color}

\newcommand{\ben}{\begin{enumerate}}
\newcommand{\een}{\end{enumerate}}
\newcommand{\bea}{\begin{eqnarray}}
\newcommand{\eea}{\end{eqnarray}}
\newcommand{\be}{\begin{equation}}
\newcommand{\ee}{\end{equation}}
\def\mv{m_{\varphi}}

\usepackage{scrextend}
\begin{document}

\preprint{APS/123-QED}

\title{Non-thermal warm dark matter limits from small-scale structure}

\author{Arka Banerjee}
\address{Department of Physics, Indian Institute of Science Education and Research, Homi Bhabha Road, Pashan, Pune 411008, India}

\author{Subinoy Das}
\address{Indian Institute of Astrophysics, Bengaluru, Karnataka 560034, India}

\author{Anshuman Maharana}
\address{Harish-Chandra Research Institute, A CI of Homi Bhabha National Institute, Chhatnag Road, Jhunsi, Prayagraj, Uttar Pradesh 211019, India}

\author{Ethan O.~Nadler}
\address{Carnegie Observatories, 813 Santa Barbara Street, Pasadena, CA 91101, USA}
\address{Department of Physics $\&$ Astronomy, University of Southern California, Los Angeles, CA, 90007, USA}

\author{Ravi Kumar Sharma}
\email{ravi.sharma@iiap.res.in}
\address{Indian Institute of Astrophysics, Bengaluru, Karnataka 560034, India}

\begin{abstract}
We present small-scale structure constraints on sterile dark matter produced from a heavy mediator particle, inspired by models of moduli decay. Dark matter particles produced through this mechanism can contribute to the entire dark matter energy density but the particles  have a non-thermal phase-space distribution; however, we show that the resulting linear matter power spectra can be mapped to effective thermal-relic warm dark matter models. This production mechanism is therefore subject to warm dark matter constraints from small-scale structure as probed by ultra-faint dwarf galaxy abundances and strong gravitational lensing flux ratio statistics. We use the correspondence to thermal-relic models to derive a lower bound on the non-thermal particle mass of $107\ \mathrm{keV}$, at $95\%$ confidence. These are the first and most stringent constraints derived on  
sterile dark matter produced via the heavy mediator decay scenario we consider. 
\end{abstract}

\maketitle


\section{Introduction} 
\label{sec:intro}

The concordance $\Lambda$CDM model has been successful in describing the accelerated expansion of the Universe, as well as the evolution of perturbations on large scales, and fits most cosmological observations to date. Observations, including those from the Cosmic Microwave Background, galaxy clustering, weak lensing, and Lyman-$\alpha$ measurements imply that matter accounts for nearly 30 percent of the total energy density of the Universe today~\cite{BOSS:2016wmc,Planck:2018vyg,DES:2021wwk}. Most of this matter is \emph{dark}; in the $\Lambda$CDM model, the dark matter is cold, meaning that its free-streaming effects are negligible. Although CDM is the simplest dark matter model that describes the data, potential tensions on small, non-linear scales have been studied for several decades \cite{Bullock:2017xww}. Historically, the ``missing satellites'' \cite{Klypin:1999uc,Moore:1999nt}, ``core--cusp'' \cite{Flores:1994gz,Moore:1994yx}, and ``too big to fail'' \cite{2011MNRAS.415L..40B} problems have received the most attention, while subtler measurements of galaxy diversity have been discussed in recent years~\cite{Sales:2022ich}.

Although most---if not all---of these potential tensions can be alleviated by baryonic physics, they have also inspired proposals for dark matter physics beyond the CDM paradigm. One classic model is that of warm dark matter (WDM), in which dark matter particles free stream on macroscopic scales of $\mathcal{O}(\mathrm{kpc})$ or larger, imprinting an observable suppression of small-scale structure relative to CDM \cite{Bond:1983hb,Bode:2000gq}. WDM particle candidates, including sterile neutrinos, are often modeled with production mechanisms that yield a thermal phase-space distribution \cite{Dodelson:1993je}, which is strongly constrained by small-scale structure \cite{Abazajian:2006yn}. Several production mechanisms that yield non-thermal distributions have also been proposed~\cite{Shi:1998km,DeGouvea:2019wpf}, although structure formation constraints on these models are also stringent~\cite{DES:2020fxi,Zelko:2022tgf,An:2023mkf}. 

Here, we study the impact of WDM production via the decay of a heavy scalar particle, inspired by models of moduli decay \cite{Bhattacharya:2020zap,Das:2021pof,Banerjee:2022era} where hot dark matter is produced from the moduli decay in early universe and contributes to a tiny fraction of total DM budget. Here, in this work,  we instead address the scenario where the entire dark matter relic density is produced in a warm state via heavy scalar decay. We show that, despite its non-thermal phase-space distribution, this model yields linear matter power spectra that are indistinguishable from thermal-relic WDM but with a non-trivial mapping to a thermal-relic mass; this correspondence allows us to derive constraints on the non-thermal particle mass from state-of-the-art small-scale structure observations.

The small-scale structure limits we leverage were derived from Dark Energy Survey \cite{DES:2005dhi} and Pan-STARRS1~\cite{Chambers:2016jzn} observations of the Milky Way satellite galaxy population \cite{DES:2019vzn}. The abundance of observed satellites, and particularly the smallest ``ultra-faint'' dwarf galaxies detected nearby, has been used to place a lower limit on the abundance of low-mass~($\gtrsim 10^8~M_{\mathrm{\odot}}$) subhalos orbiting the Milky Way \cite{DES:2019ltu}. In turn, this measurement constrains any dark matter physics that reduces the abundance of such subhalos and yields a lower limit on the thermal-relic WDM particle mass of $6.5~\mathrm{keV}$ at $95\%$ confidence \cite{DES:2020fxi,Maamari:2020aqz,Das:2020nwc,Nguyen:2021cnb,DES:2022doi,An:2023mkf}. We also leverage constraints from strong gravitational lensing. In particular, observations of flux ratio statistics of quadruply-imaged quasars lensed by low-redshift elliptical galaxies have been used to derive constraints on WDM \cite{Gilman:2019bdm} and other dark matter physics~\cite{Gilman:2022ida,Laroche:2022pjm,Dike:2022heo}. Furthermore, Ref.~\cite{Nadler:2021dft} combined WDM constraints from strong lensing and Milky Way satellite galaxies to derive a lower limit on the thermal-relic mass to date of $9.7~\mathrm{keV}$ at $95\%$ confidence, which we will use to derive the most stringent limit on our non-thermal production mechanism. 

Importantly, both thermal-relic limits WDM we use, from Milky Way satellites alone and their combination with strong lensing, are marginalized over uncertainties in the properties of the respective systems and other non-linear effects (e.g., galaxy formation physics in the satellite case and tidal stripping in the lensing case) that may affect the observables, ensuring that the constraints we derive here are conservative. Furthermore, because our model yields transfer functions that are nearly identical to thermal-relic WDM and does not introduce additional non-linear physics, the constraints we derive by mapping to existing WDM limits are robust.


\section{Production Mechanism and Phase-space Distribution} 
\label{sec:Production}

WDM scenarios are characterized by the primordial phase-space distribution of the warm species, which is determined by the production mechanism. We consider WDM produced from the decay of a heavy scalar, following Refs.~\cite{1212.4160, 1304.1804, 1908.10369, 2020zap}. 
In this scenario, the energy density of the universe is dominated by a cold, heavy species $\varphi$ at early times;
this epoch can be caused by perturbative reheating after inflation, or by moduli domination. $\varphi$ then decays to the 
Standard Model sector and WDM particles. The decay channel to WDM particles is $1 \to 2$ with the production of two identical relativistic WDM particles. While the Standard Model sector thermalizes, the WDM particles are assumed to be sterile and thus do not thermalize. The  model is natural to consider if one is interested in sterile dark matter produced  from the decay of a heavy parent
Furthermore, it would also arise in cases where sterile dark matter is produced directly from the decay of the inflaton.

\begin{figure}[t!]
\centering
\includegraphics[width=0.475\textwidth]{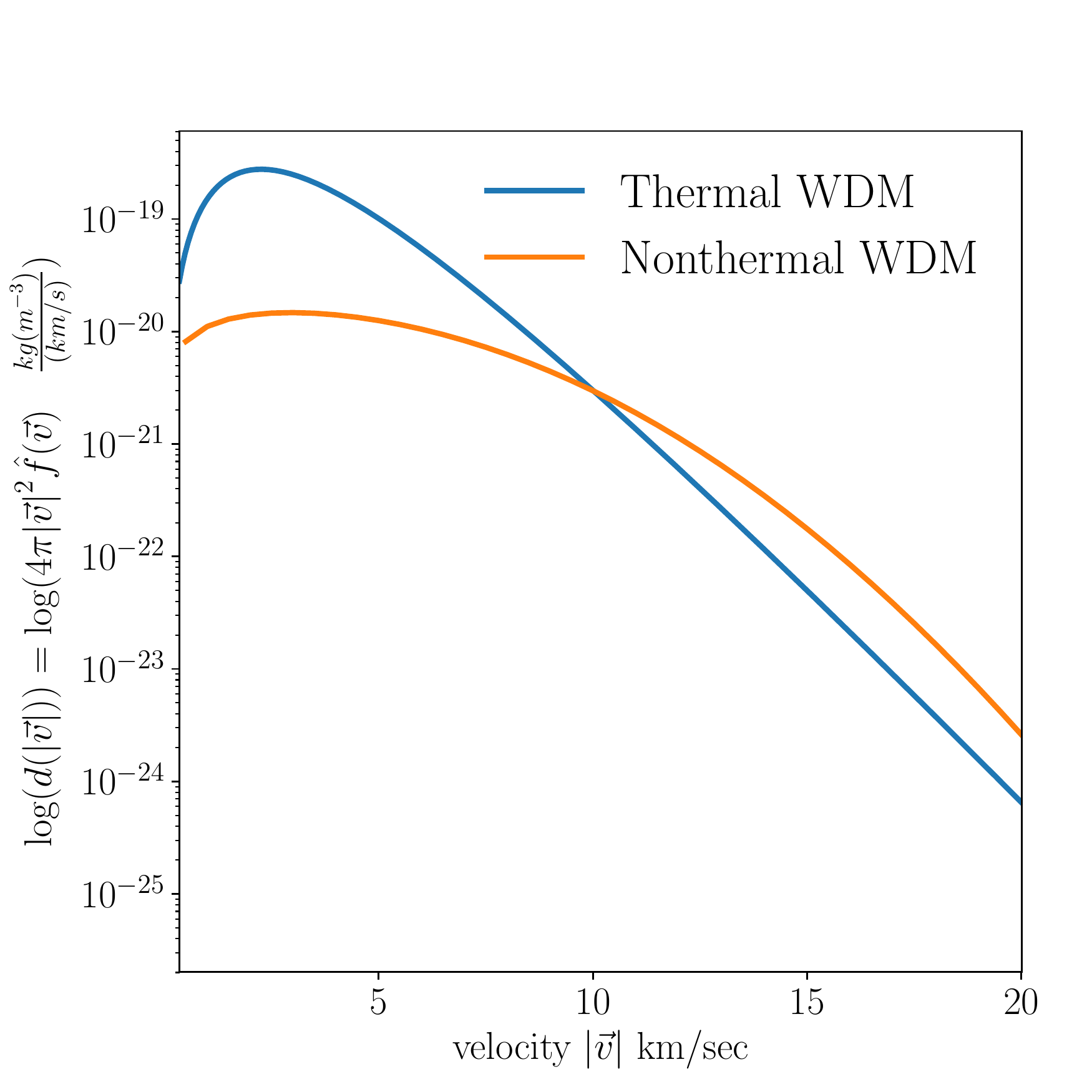}

\caption{Velocity distributions for thermal-relic WDM (blue) and our non-thermal WDM model (orange) at $z=99$ for equal masses of $5\ \mathrm{keV}$.}
\label{compareTh}
\end{figure}

Thus, the WDM production rate is determined by the decay rate of $\varphi$ and the branching ratio of the WDM channel. As $\varphi$ decays,
WDM particles are constantly produced, with momentum $ m_{\varphi}/2$ at the time of production. The late-time momentum of a WDM particle is determined by redshifting its initial momentum. Because different WDM particles are produced at different times (but with the same initial momentum), this
redshifting effect leads to a non-thermal momentum distribution, as shown in Fig.~\ref{compareTh}.

\begin{figure*}[t!]
\includegraphics[width=0.475\textwidth]{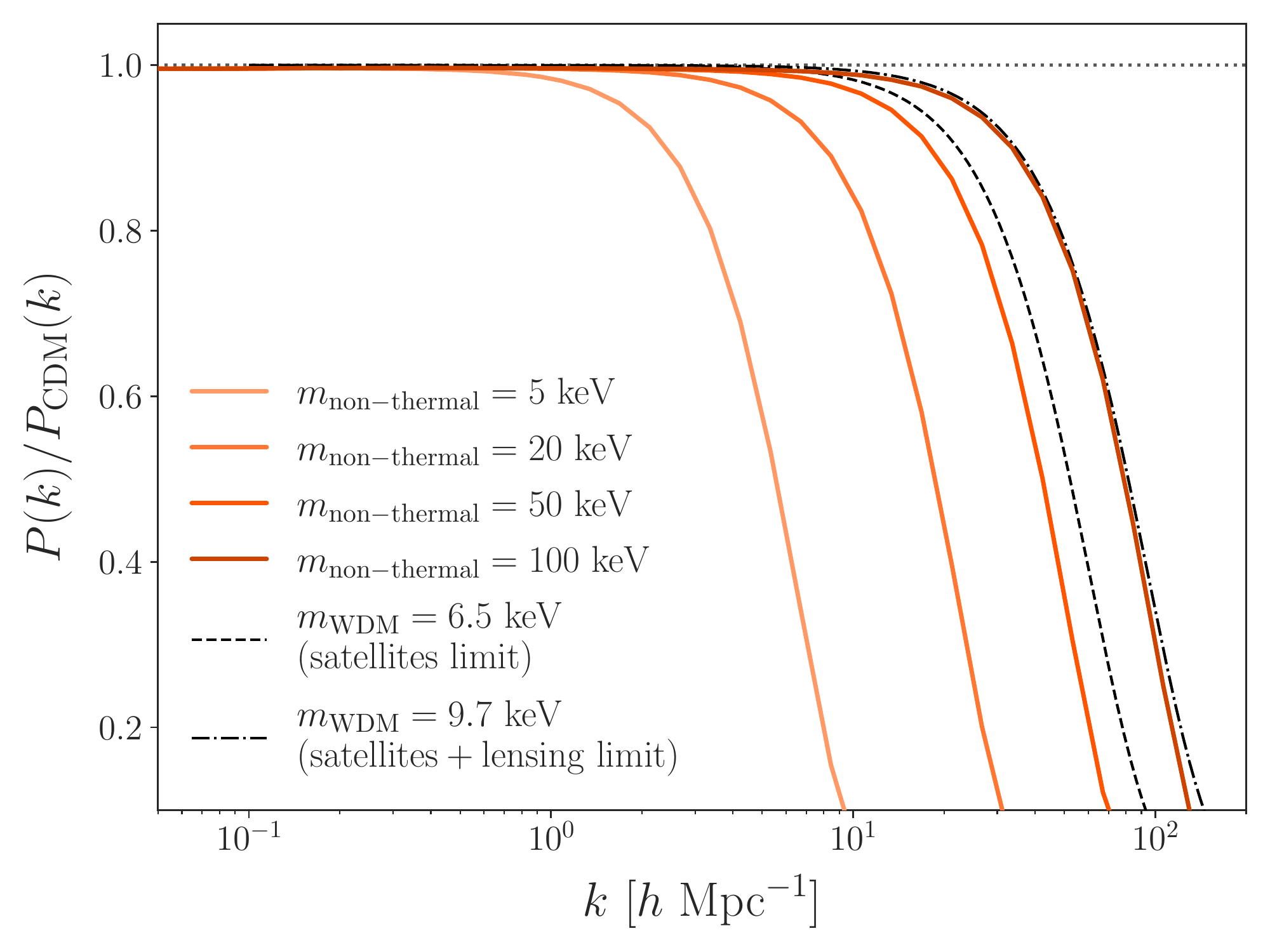}
\includegraphics[width=0.475\textwidth]{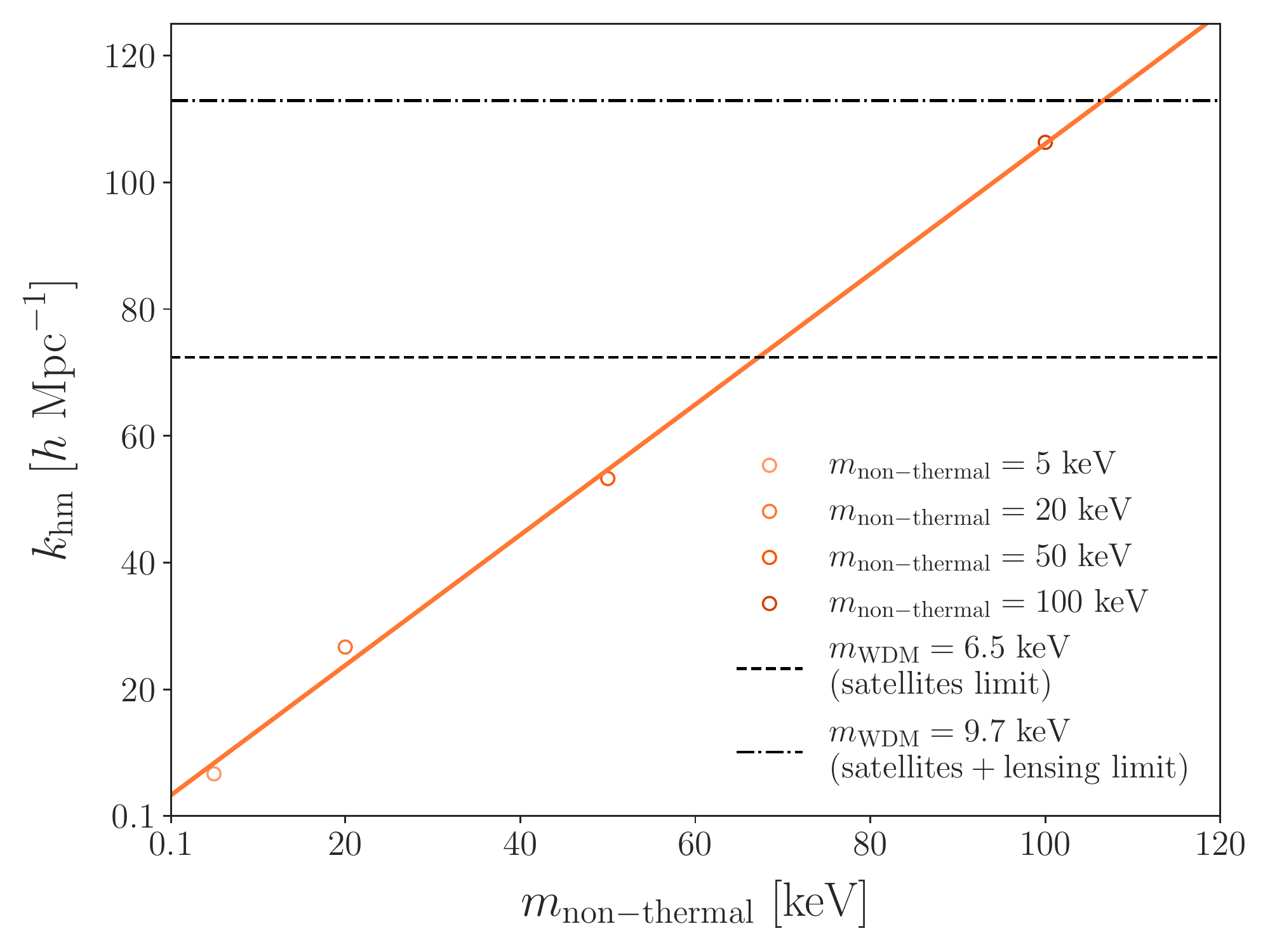}
\caption{\emph{Left}: Ratio of the linear matter power spectrum relative to CDM for our non-thermal warm dark matter model (red lines, with masses of $5\ \mathrm{keV}$ to $100\ \mathrm{keV}$, from lightest to darkest colors). Black dashed and dot-dashed lines respectively show WDM transfer functions for $6.5\ \mathrm{keV}$ and $9.7\ \mathrm{keV}$ thermal relics, which are respectively ruled out at $95\%$ confidence by analyses of the Milky Way satellite galaxy population \cite{DES:2020fxi} and its combination with strong lensing flux ratio statistics \cite{Nadler:2021dft}. \emph{Right}: The half-mode wavenumber, at which the linear matter power spectrum drops to $25\%$ that in CDM, versus non-thermal warm dark matter particle mass. The solid red line shows the fit to our \texttt{CLASS} output given by Eq.~\ref{eq:khm}; open circles show half-mode wavenumbers derived from our \texttt{CLASS} output for the non-thermal models shown in the left panel. For reference, horizontal dashed (dot-dashed) lines show half-mode wavenumbers for $6.5\ \mathrm{keV}$ ($9.7\ \mathrm{keV}$) thermal-relic WDM models, respectively.}
\label{fig:khm_relation}
\end{figure*}

The distribution function in our model is characterized by the mass of the heavy particle, $(m_{\varphi})$, its decay rate, $\tau$, and the branching ratio for decay to the WDM, $B_{\rm sp}$. In particular, the momentum distribution of the WDM particles can be computed from the fact that, as a result of the decays, the co-moving number density of $\varphi$ falls off as $N(t) = N(0) e^{-t/ \tau}$ with a branching ratio $B_{\rm sp}$ to the WDM particles. Once produced, the WDM particles free-stream. 
The distribution function today was obtained in Ref.~\cite{2020zap} as
 \be
 \label{fq}
   g(\vec{q}) =  {  32 T^{3}_{\rm ncdm, 0}\over { \pi \hat{E}^3 } } \left( { N(0) B_{\rm sp}  \over { \hat{s}^{3}(\theta^{*} ) } } \right)
     { e^{- \hat{s}^{-1}(y)} \over |\vec{q}|^{3} \hat{H}(\hat{s}^{-1}(y))}.
 \ee
Here, $T_{\rm ncdm,0}$ is the typical magnitude of the  momentum of the particles and is given by
\be
\label{tmomf}
  T_{\rm ncdm,0}  =  0.418 \left( \mv^2 \tau \over M_{\rm pl} \right)^{1/2} { T_{\rm cmb}  \over {(1 - B_{\rm sp})^{1/4} }} \equiv \zeta T_{\rm cmb}.
\ee
In Eq.~\ref{fq}, $\hat{E} = m_{\varphi}/2$ is the energy of a WDM particle at the time of its production, $N(0)$ is the initial number density  of $\varphi$ particles, and $\hat{s}(\theta)$ is the scale factor as a function of the dimensionless time $\theta \equiv {t / \tau}$. The scale factor convention in 
Eq.~\ref{fq} is such that  $\hat{s}=1$ at $\theta = 0$. Next, $\hat{s}(\theta^*)$ is its value at a fiducial value $\theta^{*}$, at which most of the $\varphi$ particles would have decayed (we take $\theta^{*} = 15$ for our computations), $\hat{s}^{-1}$ is the functional inverse of the 
scale factor as a function of the dimensionless time, and $\hat{H} = \hat{s}'(\theta) /  \hat{s}(\theta)$ is the dimensionless Hubble parameter. Finally, $y$ is defined as 
\be
  \label{yyyfd}
     y = { |\vec{q}| \over 4} \hat{s}(\theta^*),
\ee
where $\vec{q}$ is the momentum in units of the typical momentum magnitude of the WDM particles today, $T_{\rm ncdm,0}$. Note that $\vec{q}$ is constrained such that
\be
\label{qran}
   {4 \over \hat{s}(\theta^{*}) } < |\vec{q}| < 4.
\ee

We use the publicly available package \texttt{CLASS} \cite{class101, class102} to study the effects of WDM in the above distribution. For this purpose, it is useful to consider the distribution function in  Eq.~\ref{fq} divided by $T^3_{\rm ncdm, 0}$, i.e.
 \be
 \label{fq1}
   f(\vec{q}) =  {  32 \over { \pi \hat{E}^3 } } \left( { N(0) B_{\rm sp}  \over { \hat{s}^{3}(\theta^{*} ) } } \right)
     { e^{- \hat{s}^{-1}(y)} \over |\vec{q}|^{3} \hat{H}(\hat{s}^{-1}(y))}.
 \ee

We compare a typical momentum distribution in our model to that of thermal-relic WDM in Fig.~\ref{compareTh}. For the same value of $\Delta N_{\rm eff}$ (the effective number of additional neutrino-like species) at the time of Big Bang Nucleosynthesis (BBN), the non-thermal distribution is 
peaked at higher values of the momentum and is much broader than the thermal distribution. Note that the small-scale structure constraints we derive below imply that the allowed masses of WDM particles in our model is well below the temperature of the universe during BBN. 
 
Note that although naively $f(\vec{q})$ seems to depend on $N(0)$, the full expression is independent of $N(0)$ as long as we take the universe to be completely matter-dominated at the initial time \cite{2020zap}. On the other hand, changing the parameter $B_{\rm sp}$ scales the distribution function by an overall constant. For any value of the mass of the WDM particle, the \texttt{CLASS} package scales the distribution function so that $\Omega_{\rm WDM}$ remains consistent with observations. Thus, values of $B_{\rm sp}$ are effectively sampled in our analysis. The other parameters in the distribution function are $m_{\varphi}$ and $\tau$. Motivated by high-scale inflation as the theory of the early universe and the decay of the inflaton by perturbative GUT scale interactions, our interest is in $m_{\varphi}\sim {\mathcal{O}}(10^{-6} M_{\rm pl})$ and $\tau \sim {\mathcal{O}}(10^{8} / m_{\varphi})$. In the Appendix, we show that our results are insensitive to the precise values of these parameters.
 

\section{Linear Matter Power Spectrum and Transfer Function}
\label{sec:Pk}

We use a modified version of \texttt{CLASS}\footnote{https://github.com/ravi398/WDM} \cite{class101,class102}  to generate linear matter power spectra for our non-thermal warm dark matter model. 
Throughout, we adopt the same cosmological parameters as Ref.~\cite{Das:2021pof}. For our WDM CLASS runs, we set $\omega_{\mathrm{cdm}}=0$. We include two non-CDM species, the first of which represents standard massless neutrinos, and the second of which is the WDM species with $N_{\mathrm{ncdm}}=2$ and $\omega_{\mathrm{ncdm}}=0.12$. We fix the fiducial phase-space distribution parameters for our non-thermal WDM production mechanism following Ref.~\cite{Das:2021pof}, with $B_{\mathrm{sp}}=0.0118$, $m_{\phi}=10^{-6}M_{pl}$, and $\tau=10^8/m_{\phi}$. We show in the Appendix that our results are not sensitive to the choices of these parameters; thus, our constraints are not sensitive to this choice.

The left panel of Fig.~\ref{fig:khm_relation} shows the ratio of the linear matter power spectrum relative to CDM, hereafter referred to as the \emph{transfer function}, for various non-thermal particle masses. In particular, we define
\begin{equation}
    T^2(k) \equiv \frac{P(k)}{P_{\mathrm{CDM}}(k)}.
\end{equation}
We compare these transfer functions to the thermal-relic warm dark matter transfer function fit from Ref.~\cite{Viel:2005qj}. Note that the relation between thermal-relic mass and cutoff scale derived from the Ref.~\cite{Viel:2005qj} fitting function is inaccurate for sufficiently cold models, including those near our $95\%$ confidence constraints; however, the shape of Ref.~\cite{Viel:2005qj} transfer function cutoff is accurate \cite{Decant:2021mhj,Vogel:2022odl}. Because the analyses we compare to also assumed the Ref.~\cite{Viel:2005qj} functional form, which we only use as a means to map to effective thermal-relic models, this discrepancy does not affect our constraints.

The shape of the transfer function cutoff  in our non-thermal model is clearly similar to that in thermal-relic warm dark matter. This allows us to construct a mapping between the models, following methodology similar to that in Refs.~\cite{Nadler:2019zrb,Nadler:2021dft}. As illustrated in Fig.~\ref{fig:khm_relation}, we find that (for our fiducial cosmological parameters) the half-mode scale $k_{\mathrm{hm}}$, defined by $T^2(k_{\mathrm{hm}})\equiv 0.25$, is related to the non-thermal particle mass via
\begin{equation}
    k_{\mathrm{hm}} = \left(3.3 + \frac{m_{\mathrm{non-thermal}}}{\mathrm{keV}} \right)\ h\ \mathrm{Mpc}^{-1},
    \label{eq:khm}
\end{equation}
where $m_{\mathrm{WDM}}$ is our non-thermal particle mass. Setting Eq.~\ref{eq:khm} equal to the thermal-relic warm dark matter half-mode scale \cite{Viel:2005qj} yields the relation
\begin{equation}
    m_{\mathrm{non-thermal}} = 12.2\left(-0.9+\frac{m_{\mathrm{WDM}}}{\mathrm{keV}} \right)\ \mathrm{keV},
    \label{eq:mapping}
\end{equation}
where $m_{\mathrm{WDM}}$ is the thermal-relic WDM mass. We have verified that this relation is accurate over the entire range of non-thermal masses we consider by comparing the corresponding transfer functions. Along the initial cutoff, discrepancies between matched transfer functions are typically at the sub-percent level; such deviations are not detectable in the datasets on which we base our constraints. Thus, Eq.~\ref{eq:mapping} provides a means to robustly translate existing thermal-relic warm dark matter constraints to constraints on our non-thermal model.

Note that $m_{\mathrm{non-thermal}}$ is significantly larger than the corresponding $m_{\mathrm{WDM}}$ based on Eq.~\ref{eq:mapping}. 
A similar result was found
for hot dark matter in the non-thermal distribution under consideration \cite{2020zap}. This essentially follows from the fact that the non-thermal momentum distribution has a long tail toward high velocities (e.g., see Figure~\ref{compareTh}). Thus, at a given redshift, non-thermal WDM particles in our model have a higher average velocity compared to thermal WDM particles, even for the same mass. 

We can gain further insight into the relation between $m_{\mathrm{non-thermal}}$ and $m_{\mathrm{WDM}}$ by calculating free-streaming length via
\begin{equation}
    \lambda_{\mathrm{fs}}=\int_{0}^{a_{\mathrm{eq}}}  \frac{\langle v(a)\rangle}{a^2 H(a)} \mathrm{d}a,
\end{equation}
which respectively has contributions from the relativistic and non-relativistic regimes given by
\begin{equation*}
    \lambda_{\mathrm{fs}}=2c t_{0} a_{nr}+ I.
\end{equation*}
Here, the non-relativistic contribution $I$ is given by 
\begin{equation*}
I= \int_{a_{\mathrm{nr}}}^{a_{\mathrm{eq}}}  \frac{\langle v(a)\rangle}{a^2 H(a)} \mathrm{d}a,
\end{equation*}
where the average velocity is 
\begin{equation*}
\langle v(a)\rangle=\frac{\int_{0}^{\infty} v f(v)4\pi v^2\mathrm{d}v}{\int_{0}^{\infty}  f(v)4\pi v^2\mathrm{d}v}.
\end{equation*}
For thermal-relic WDM, these integrals are analytically tractable assuming $H^2(a)=\Omega_{m} a^{-3}+\Omega_{\mathrm{rad}} a^{-3}+\Omega_{\Lambda}$, where $\Omega_{i}$ are fractional energy densities and $a$ is the scale factor.
For non-thermal WDM, we solve the integral equations numerically to compute the free-streaming scale as a function of $m_{\mathrm{non-thermal}}$. We have verified that the non-thermal and thermal-relic free-streaming scales match each other well along the relation given by Eq.~\ref{eq:mapping}.

\begin{figure}[t!]
\hspace{-3.5mm}
\includegraphics[width=0.475\textwidth]{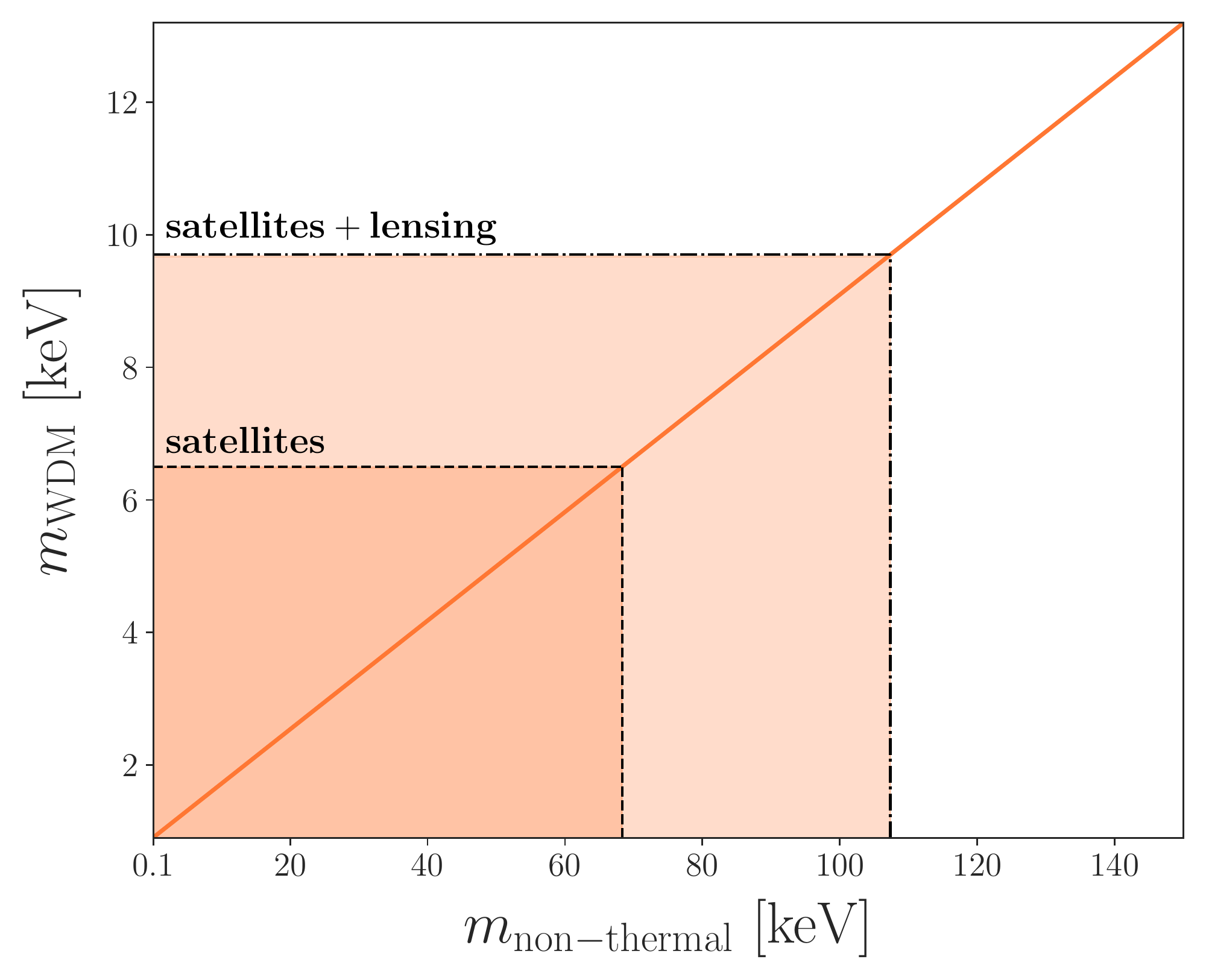}
\caption{Relation between non-thermal and thermal-relic WDM mass, and corresponding small-scale structure limits, derived by matching our non-thermal WDM linear matter power spectra to equivalent thermal-relic models. The solid red line shows the relation from Eq.~\ref{eq:mapping}; the dark and light shaded regions are respectively excluded at $95\%$ confidence by the Milky Way satellite galaxy population and its combination with strong lensing flux ratio statistics.}
\label{fig:constraints}
\end{figure}



\section{Constraints from Thermal Relic WDM Mapping}

We derive constraints on our non-thermal model based on Eq.~\ref{eq:mapping}. In particular, we use the $95\%$ confidence lower limits on thermal-relic WDM of $6.5\ \mathrm{keV}$, derived from the Milky Way satellite galaxy population, and $9.7\ \mathrm{keV}$, derived from its combination with strong lensing flux ratio statistics. These constraints are illustrated in Figure~\ref{fig:constraints} and map to lower limits on the non-thermal WDM mass of $68$ and $107\ \mathrm{keV}$, respectively. Following our calculation above, the corresponding free-streaming lengths are $\lambda_{\mathrm{fs}}=31~\mathrm{kpc}$ for $m_{\mathrm{non-thermal}}=107~\mathrm{keV}$ and $\lambda_{\mathrm{fs}}=29~\mathrm{kpc}$ for $m_{\mathrm{WDM}}=9.7~\mathrm{keV}$. The latter value is in reasonable agreement with the fit in Ref.~\cite{Schneider:2011yu}.

We emphasize that our non-thermal transfer functions are essentially identical to thermal-relic models, including for non-thermal masses at our $95\%$ confidence limits (e.g., see the left panel of Figure~\ref{fig:khm_relation}). Furthermore, there is no new, non-linear physics introduced by our model: in both the non-thermal and thermal-relic WDM cases, suppression of structure formation relative to CDM is set by the free-streaming scale, which is imprinted well before matter--radiation equality. Thus, we expect nonlinear observables that drive the small-scale structure limits we map to, such as the halo and subhalo mass functions and mass--concentration relations, to be practically identical to those in thermal-relic WDM, which are well studied. 

We note that the precise mapping between our non-thermal transfer functions and thermal-relic WDM at and below the cutoff scale justifies our use of combined Milky Way satellite and strong lensing limits. In particular, the lensing limits depend sensitively on halo concentrations \citep{Gilman:2019bdm,Gilman:2021gkj}, which are in turn sensitive to power on scales even smaller than $k_{\mathrm{hm}}$.

\section{Conclusions}

We have derived the first small-scale structure limits on WDM produced via heavy scalar decay. This production mechanism is theoretically motivated and yields a distinctly non-thermal primordial phase-space distribution. Nonetheless, assuming that the entire dark matter relic density is composed of these warm particles, we have shown that the resulting linear matter power spectra are nearly identical to thermal-relic WDM. We have exploited this correspondence to derive a mapping between our non-thermal particle mass and the standard thermal-relic WDM mass. Confronting these small-scale power spectra with observational constraints from Milky Way satellite galaxies and strong gravitational lensing yields a lower limit on the non-thermal particle mass of $107~\mathrm{keV}$ at $95\%$ confidence.

The sensitivity of both small-scale structure probes we leveraged will improve significantly over the next decade. Specifically, the combination of next-generation observational facilities, including the Vera C.\ Rubin Observatory \cite{LSST:2008ijt}, and space telescopes, including the Nancy Grace Roman Space Telescope \cite{Spergel:2015sza}, is expected to probe most of the undiscovered Milky Way satellite galaxy population as well as faint dwarf galaxies throughout the Local Volume \cite{LSSTDarkMatterGroup:2019mwo,Gezari:2022rml}. These same facilities will drastically increase the sample of strongly-lensed systems available for deep follow-up observations necessary to derive dark matter constraints \cite{Oguri:2010ns,Weiner201015173}, and will also provide qualitatively new means of probing low-mass halos \cite{Pardo:2021uzy}. Together, these observations are expected to probe warm dark matter masses up to $\sim 20\ \mathrm{keV}$ \cite{LSSTDarkMatterGroup:2019mwo}, which translate to non-thermal particles masses of $\sim 200\ \mathrm{keV}$ for our production mechanism. 

The forecasts above assume that warm dark matter particles compose the entire dark sector. Smaller fractions of lower-mass warm dark matter particles may evade next-generation constraints. Here, we have essentially treated the parameters in our non-thermal distribution functions phenomenologically; however, it will therefore be interesting to consider explicit realizations of this scenario that naturally yield mixed cold-plus-warm dark matter scenarios. In the context of string theory, two WDM production scenarios are particularly relevant: the decay of the inflaton to excitations at the bottom of a warped  throat \cite{Frey:2009qb}, or a hidden sector with keV-scale condensation (e.g., see Ref.~\cite{Halverson:2018vbo}). A generic outcome of such string models is that dark matter can be mixed, with one component arising from the decay of a heavy modulus (as discussed in this paper) and the other corresponding to a thermal relic that freezes out from the Standard Model sector (see Ref.~\cite{stringR} for a recent review of various scenarios for dark matter in string theory). We plan to pursue detailed studies of these possibilities in future work.


\appendix

\section{Variations in Non-thermal Production Mechanism Parameters}
\label{sec:production_variations}

\begin{figure}[ht!]
    \centering
    \includegraphics[width=0.5 \textwidth]{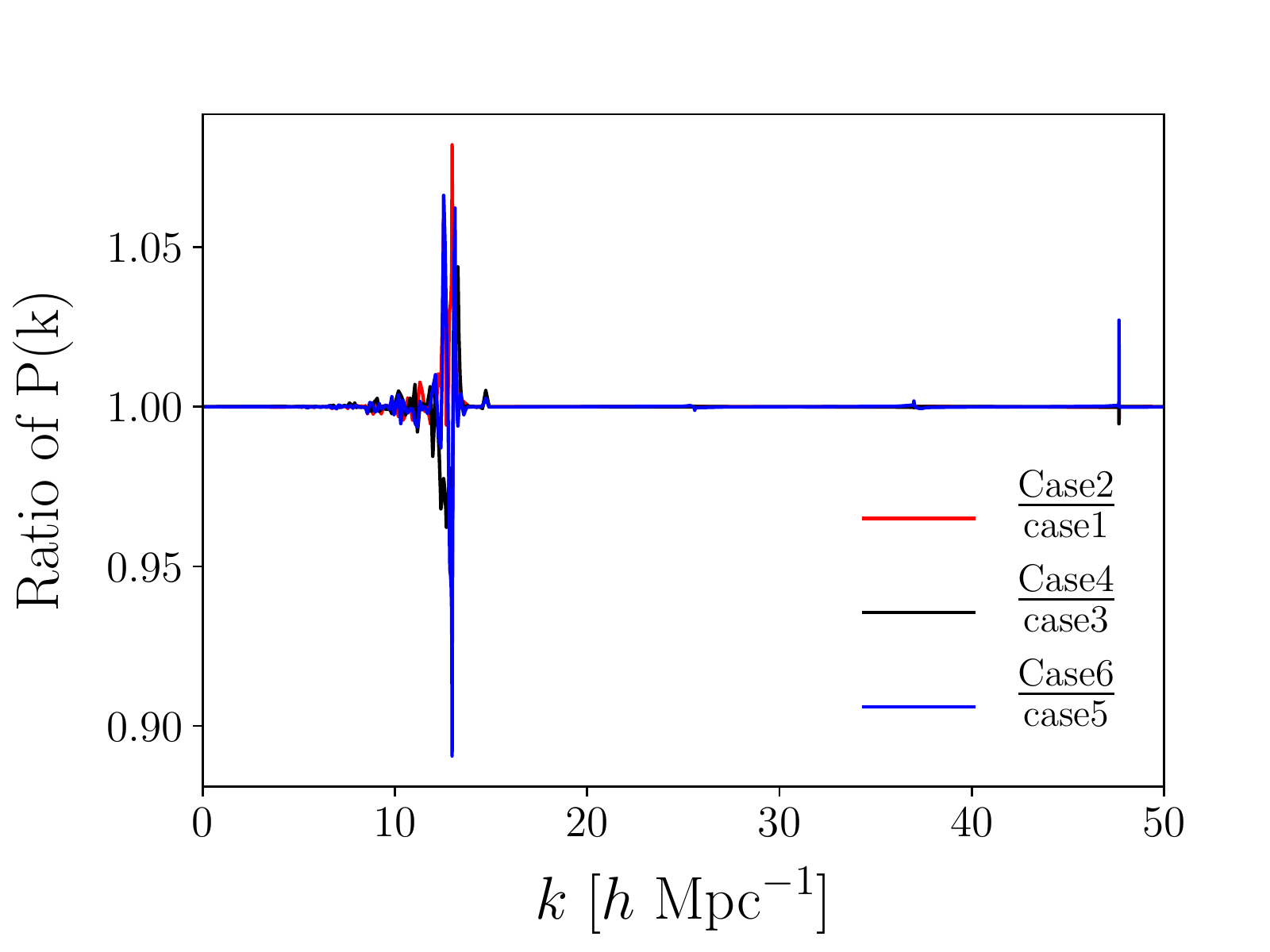}
    \caption{Ratio of the linear matter power spectrum for sets of production mechanism parameter variations, as listed in Table~\ref{table:variation}. The kinks that appear around $k=12 h\ \mathrm{Mpc}^{-1}$ are the result of small numerical artifacts in CLASS code; their location coincides with oscillations in non-thermal WDM power spectra at scales smaller than the initial cutoff.}
    \label{fig:variation_pk}
\end{figure}

As discussed above, the distribution function for our non-thermal model is characterized by the mass of the heavy particle, $m_{\phi}$, its decay rate, $\tau$, and
the branching ratio for decay to WDM, $B_{\rm sp}$. Here, we explicitly check that variations in these parameters do not strongly impact the linear matter power spectra the we use to construct a mapping to thermal-relic WDM. 
Specifically, we choose a set of variations in the as listed in Table~\ref{table:variation}. The ratio of the linear matter power spectrum between parameter sets is shown in Fig.~\ref{fig:variation_pk}. We can see that power spectra are not strongly affected by these parameter variations, implying that our results are robust to specific choices regarding the details of our non-thermal WDM production mechanism.

\begin{table*}[t!]
    \centering
   
\begin{tabular}{ |c|c|c|c|c|c|c| } 

 \hline
  Model Parameters & case 1 & case 2 & case3 & case 4 &case5 &case 6 \\ 
 \hline
 $m_{\varphi} $& $10^{-6} M_{\rm pl}$ & $10^{-8} M_{\rm pl}$& $10^{-7} M_{\rm pl}$ & $10^{-8} M_{\rm pl}$& $10^{-6} M_{\rm pl}$& $10^{-8} M_{\rm pl}$ \\ 
 \hline
 $\tau$ & $10^{8} \big{/} m_{\varphi}$ & $10^{9} \big{/} m_{\rm \varphi}$ & $10^{8} \big{/} m_{\varphi}$ & $10^{8} \big{/} m_{\rm \varphi}$& $10^{8} \big{/} m_{\rm \varphi}$& $10^{9} \big{/} m_{\rm \varphi}$\\ 
 \hline
 $m_{\rm sp}$ & 5000 &  5000 & 5000 &   5000& 5000 &   5000\\
 \hline
 $B_{\rm sp}$ & 0.0118 & 0.0118 & 0.0118 & 0.0118& 0.0218 & 0.0218 \\
 \hline
\end{tabular}
\label{table:variation}
\caption{Variations of production mechanism parameters used for the test in Fig.~\ref{fig:variation_pk}. Note that $m_{\rm sp}$ is given in eV.}
\end{table*}

\pagebreak
\bibliography{apssamp}

\end{document}